\documentclass[12pt]{iopart}

\usepackage{iopams}  
\newcommand{\be}{\begin{equation}}
\newcommand{\ee}{\end{equation}}
\newcommand{\bea}{\begin{eqnarray}}
\newcommand{\eea}{\end{eqnarray}}
\usepackage{float}
\usepackage{graphicx}
\usepackage{xcolor}
\usepackage{bigints}
\usepackage{amsmath}
\usepackage{amsmath}
\DeclareMathOperator{\sech}{sech}

\begin{document}

\title{Exact Solutions of Augmented GP Equation: Solitons, Droplets and Supersolid}

\author{Subhojit Pal }
\address{Department of Physics, Indian Institute of Science Education and Research (IISER), Bhopal, Madhya Pradesh 462066, India}
\ead{subhojit18@iiserb.ac.in}

\author{Aradhya Shukla}
\address{ Department of Physics, Institute of Applied Science and Humanities, 
		GLA University,\\ Mathura 281406, India}%
\ead{ashukla038@gmail.com}

\author{ Prasanta K. Panigrahi}
 \address{ Department of Physical Sciences, Indian Institute of Science Education and Research Kolkata, Mohanpur-741246, West Bengal, India}%
\ead{pprasanta@iiserkol.ac.in}
\date{\today}
\begin{abstract}
The augmented nonlinear Schr\"odinger equation (ANLSE), describing BEC, with the  Lee-Huang-Yang (LHY) correction has exhibited a quantum droplet state, which has found experimental verification. In addition to the droplet, exact kink-antikink  and supersolid phases have been recently obtained in different parameter domains. Interestingly, these solutions are associated with a constant background, unlike the form of BEC in quasi-one dimension, where dark, bright, and grey solitons have been experimentally obtained. Here, we connect a wide class of solutions of the ANLSE with the Jacobi elliptic functions using a fractional transformation method in a general scenario. The conserved energy and momentum are obtained in this general setting which differentiates and characterizes the different phases of the solution space. We then concentrate on the Jacobi-elliptic   $dn(x, m^2)$ function, as the same is characterized by a non-vanishing background as compared to the other $cn$ and $sn$ functions.
\end{abstract}

\maketitle

\section{Introduction}
The existence of solitons in non-linear systems are well established \cite{drezin} and has been observed in various physical systems \cite{sol, bis, poly_acy}, e.g.,  polyacetelene \cite{pkp_poly}, optical fibers\cite{has, GPagr} and Bose-Einstein condensates (BECs) to mention a few \cite{pethick, kev}. It is worth mentioning that depending upon their physical properties,  the solitons are of the forms of dark \cite{densch,dark, busch}, grey \cite{sh} and bright \cite{kh, mcd, aspect}. The stability of solitons is due the balancing effect between dispersion and non-linearity of the system. The grey soliton emerges in the second-quantized model,  representing the well-known Lieb mode \cite{lieb}. The grey and dark solitons are exact solutions of the non-linear Schr{\" o}dinger equation (NLSE) \cite{Kul, 1r27, jack} for the range of repulsive non-linearity, having a functional form of tangent whereas, the bright soliton is represented by the form of secant, which vanishes at the spatial asymptotic ends manifesting in the attractive scenario \cite{kiv}. These solitons also manifest in optical fibers, in the regime of repulsive and attractive Kerr non-linearities, respectively \cite{GPagr, kiv}.

In the weak coupling limit, the BEC is described by the mean field (MF) Gross–Pitaevskii (GP) equation, and various solitonic solutions in BEC has been well explained \cite{busch, sh, kh}. Interestingly, a new type of quantum matter in BEC has been observed recently, known as quantum droplet \cite{1r5, 1r5_ptr}, where the beyond mean field (BMF) Lee-Huang-Yang (LHY) correction plays a crucial role \cite{LUY1,LUY2}. It is to be noted that the BMF correction yields the MF equation, having cubic-quadratic non-linearities \cite{astra}. The quantum droplet emerges due to the balancing effect of MF and BMF corrections. It is worth mentioning that the BMF effect originates from the summation of zero point energy of Bogoliubov modes \cite{1r5, 1r5_ptr, 1r16} and it depends upon the dimensions of the system. The BMF correction in three dimensions is repulsive and varies as $n^{5/2}$ where $n$ represents the number density; on the other hand,  in one dimension it is attractive and scales as $n^{3/2}$ \cite{1r16, malo}. The shape and size of the quantum droplets depend upon the number density e.g., the flat-top shape of the droplet is due to the uniform number density \cite{barbut}.  The self-bound quantum droplet has been experimentally observed in Bose-Bose (BB) mixture \cite{ 1r16, 1r14, 1r15} as well as dipolar BEC \cite{3r1} in agreement with experimental observations \cite{FSM1, FSM2}. Recently, an exact solution for droplets emerging from the competing MF and BMF corrections has been established \cite{1r5_ptr},  which is characteristic of flat-top nature. A pair of kink/anti-kink\cite{pkp_new} solution having a constant background is obtained in the MF repulsive domain with BMF correction \cite{pkp_new}, where the lowest value of chemical potential is same on that of the self-trapped droplets \cite{ 1r5_ptr}.

The dynamics of $1-D$ quantum droplet has been explored by observing two different domains where, one of the domains exhibits the shape of quantum droplet as Gaussian and in other domain has the puddle shape \cite{astra}. In BB mixture, the existence of two types of macroscopic self-bound phases: soliton and droplet, have been demonstrated \cite{1r14}. It has also been found that droplets and solitons, depend upon the number density and scattering length, can co-exist as well as smoothly connected. It is worth mentioning that the nature of BMF correction depends upon dimensions of the system. The different behaviour of BMF have been studied in the dimensional cross-over $1D \longrightarrow 3D$ \cite{Lavoine} and $2D \longrightarrow 3D$ \cite{malo} for binary BEC. The stability of structures emerging from NLSE e.g. soliton, quantum droplet and supersolid etc.  can be understood by analysing the Modulation instability (MI). Recently, the MI analysis has been explored with and without spin-orbit coupling (SOC) to identify the parameter domains in which the the instabilty of propagating wave exhibits the soliton solutions. It is observed that the absence of SOC yields the same analytical MI expression for both \cite{pkp, kare}. However, in SOC limit, the stability of BB mixture occurs when the perturbation frequency, $\Omega_{\pm} > 0$ and for one-dimensional quantum droplets, the stability happens for $\Omega^2>0$. Recently, supesrsolid phase for trapped droplet array has been investigated, where the spontaneous breaking of rotational symmetry yields the supersolid phase in a certain domain \cite{pfou3}.


The NLSE describing the mean field dynamics of the cigar shaped BEC is a well known integrable dynamical system. Its solution space has been extensively investigated, consisting of dark, bright, grey solitons, multi solitons and rogue waves. In comparison the augmented NLSE with the Kerr non-linearity and non-analytic LHY extension has not been systematically investigated with regard to analogous solutions. Its integrability is yet to be established.  Therefore, the phase space components arising from the nonlinear excitations, comprising of the general elliptic functions, is of great interest, as both solitonic and sinusoidal excitations will follow from these in appropriate limiting conditions. Here, we carry out a systematic investigation through a map connecting  the solution space of ANLSE to the space of elliptic functions, making use of a Mobious transform. The parameter domains of quantum droplets, dark, bright solitons, nonlinear propagating waves, kink and anti kink type excitations as well as pure sinusoidal waves, are distinctly identified. The differences and similarities with the NLSE excitations are pointed out to highlight the role and manifestation of the LHY corrections in these modes. Very interestingly, the supersolid phase with droplets maintaining their coherence due to the presence of a constant superfluid background, possible only in the ANLSE case, is shown to transit to the incoherent droplet scenario in distinct parameter domains, much similar to the experimental observations in dipolar BEC. The effect of parametric variation on the quantum droplets is explicated, which can take it to a bright solitonic configuration, similar to the `immortal solitons' observed earlier. We also point out the differences in the excitations in the presence of both cubic and quadratic nonlinearities, as compared to the pure quadratic case, originating from the quantum corrections alone.

The plan of the paper is as follows. In Sec. 2, we briefly discuss about the augmented GP equation with BMF correction, and using the non-singular oscillatory form as an ansatz solution, the consistency conditions have been obtained. We obtain the energy and momentum corresponding to various excitations in sec. 4 and 5 respectively. Finally, we conclude the manuscript with a number of directions for further investigation.

\noindent\section{Augmented Gross-Pitaevieki (GP) Equation}
It is well-known that, in one-dimension, the beyond mean field (BMF) energy corrections to BEC equation is represented by the amended GP equation as,

\begin{equation}
	i\hbar \frac{\partial \psi}{\partial t}= - \frac{\hbar^2}{2 M} \frac{\partial^2 \psi}{\partial x^2} + \delta g {\left|\psi\right|}^2 \psi- \frac{\sqrt{2M}}{\pi \hbar} g^{3/2} {\left|\psi\right|} \psi
	\label{eqn:1}
	\end{equation}
which has cubic and quadratic nonlinearities. It is to be noted that the mean field corrections also appears from an effective two component BEC, having inter and intra-particle interaction in $0\le \delta g \le g$ domain.

We take the transformation $\psi(x,t)=\phi[\alpha(x-vt)] e^{[i(kx-\omega t)]}$ leading to 
\begin{equation}
\frac{\partial \phi}{\partial t}= -v \frac{\partial \phi}{\partial X},
	\label{eqn:2}
\end{equation}
with $X = x-vt$. The second equation, coming from the real part reads as,
\begin{equation}
{\alpha}^2 \phi'' + g_1 \phi +g_3 {\phi}^2+ g_2 {\phi}^3 =0,
	\label{eqn:3}
	\end{equation}
where $v=\frac{\hbar k}{M}$, $g_1=(\frac{2 M \mu}{{\hbar}^2}-k^2)$, $g_2=-\frac{2M\delta g}{\hbar^2}$, $g_3=\frac{(2M)^{\frac{3}{2}}}{\pi \hbar^3} g^{\frac{3}{2}}$ are constants, and prime indicates differentiation with respect to x. We intend to connect the solution of eq. ($\ref{eqn:3}$) to the elliptic  
equation $f'' + b \,f+c\, f^3=0$ through a Mobius transform with one conserved quantity $E_0= f'^2/2 + \frac{b}{2}f^2+ \frac{c}{4}f^4$. 
In the following section, we replace $f(x)$ by the elliptic function of specific types to obtain the consistency conditions for which,  Eqn. {($\ref{eqn:4}$)} behaves as a proper solution to the augmented GP equation.

\noindent\section{General Solution}
The solution of augmented GP equation can be obtained through ``fractional transformation'', connecting the solutions of solvable non-linear systems, with general form,
\begin{equation}
	\phi(X)=\frac{A + Bf(X)}{1+ D f(X)},
	\label{eqn:4}
\end{equation}
where A,B and D are real constants. The elliptic equation has non-singular oscillatory solutions of the form $cn[X|m^2]$, $dn[X|m^2]$ and $nd[X|m^2]$, with $m^2$ as modulus parameter. Here, we take  $f(X)=dn[X|m^2]$ and substituting eq. $(\ref{eqn:4})$ in Eq. $(\ref{eqn:3})$, one gets four equations with resitriction $AD \neq B$. It is worth mentioning here that   A, B, D and $\alpha$ are inter-connected by the equations, 
\begin{eqnarray}
&&	2{\alpha}^2 (AD-B)+ g_1 BD^2 +g_2 B^3 +g_3 B^2 D=0  \label{eqn:5}\\
&&{\alpha}^2 (AD-B)(-m^2+2)D +g_1 (AD^2 +2BD)+ 3g_2AB^2 \\ 
&&+g_3(B^2+2ABD)=0 \label{eqn:6}\\
&&	{\alpha}^2(AD-B)(m^2-2)+g_1(2AD+B)+g_3(2AB+A^2D)\\
&&+3 g_2 A^2B =0 \label{eqn:7}\\
&&		g_1 A + 2 {\alpha}^2 (AD-B)(m^2-1)D+ g_2 A^3+ g_3 A^2=0,
	\label{eqn:8}
\end{eqnarray}
In order to obtain the solutions of the above  equations, we assume $B=\Gamma D$. which transforms  Eq. {($\ref{eqn:7}$)} and Eq. ({$\ref{eqn:6}$}) into
\begin{equation}
	{\alpha}^2(A-\Gamma)(m^2-2)+g_1(2A+\Gamma)+g_3(2A\Gamma+A^2)+3 g_2 A^2\Gamma =0
	\label{eqn:9}
\end{equation}
\begin{equation}
	{\alpha}^2 (A-\Gamma)(2-m^2) +g_1 (A +2\Gamma)+ 3g_2A{\Gamma}^2 +g_3({\Gamma}^2+2A\Gamma)=0
	\label{eqn:10}
\end{equation}
From the above equation, one can obtain the following
\begin{equation}
{\alpha}^2(A,\Gamma)=\frac{g_1+g_3(A+ \Gamma)+3 g_2A \Gamma}{2(2-m^2)} 
    \label{eqn:11}
    \end{equation}
Furthermore, with the use of equations ({$\ref{eqn:5}$}) and ({$\ref{eqn:8}$}) together with equation ({$\ref{eqn:11}$}) leads to
\begin{equation}
\frac{1}{D^2}=y=\frac{1}{2}\bigg[(2-m^2)\, \pm \, \sqrt{m^4+ \frac{2 g_2(A-\Gamma)^2}{{\alpha}^2}}\bigg].
    \label{eqn:17}
\end{equation}

On the other hand, for obtaining the value of $A$, one has to add Eq. ({$\ref{eqn:9}$}) and Eq. ({$\ref{eqn:10}$}), which can be converted into the following form,
\begin{equation}
    A^2+p A +q=0,~\text{leading to}~A(\Gamma)=\frac{1}{2}\bigg[-p\, \pm\, \sqrt{p^2- 4q}\bigg],
    \label{eqn:18}
\end{equation}
where, $p=\frac{3g_1+3g_2+4g_3\Gamma}{g_3+3g_2\Gamma}$ and $q=\frac{3g_1\Gamma+ g_3{\Gamma}^2}{g_3+3g_2\Gamma}$.  One then obtains
\begin{equation}
    g_2 {\Gamma}^3+ g_3 {\Gamma}^2+g_1 \Gamma+ {\alpha}^2 (A-\Gamma)\bigg((2-m^2)\, + \,\sqrt{m^4+ \frac{2 g_2(A-\Gamma)^2}{{\alpha}^2}}\bigg)=0.
    \label{eqn:19}
\end{equation}
As is well known, the above cubic equation leads to three solutions of $\Gamma$($=\Gamma_1, \Gamma_2, \Gamma_3$), which can take their real values in appropriate parameter domains:

\begin{equation}
    \Gamma_1=- \frac{1}{6g_2}\Bigg[-2(6Ag_2+g_3)+\frac{\big(2\, 2^{\frac{1}{3}}\,(3g_1g_2+18A^2g_2^2+12Ag_2g_3+g_3^2-3(2-m^2+m^4)g_2{\alpha}^2\big)}{\Delta}+ 2^{\frac{2}{3}} \Delta \Bigg]
    \label{eqn:21}
\end{equation}

\begin{equation}
    \Gamma_2=\frac{1}{12g_2}\Bigg[ 4(6Ag_2+g_3)+\frac{\big(2.516(3g_1g_2+18A^2g_2^2+12Ag_2g_3+g_3^2-3(2-m^2+m^4)g_2{\alpha}^2\big)}{\Delta}+  2^{\frac{2}{3}} \Delta\Bigg]
    \label{eqn:22}
\end{equation}

\begin{equation}
    \Gamma_3=\frac{1}{12g_2}\Bigg[ 4(6Ag_2+g_3)+\frac{\big(2\, 2^{\frac{1}{3}}\, (3g_1g_2+18A^2g_2^2+12Ag_2g_3+g_3^2-3(2-m^2+m^4)g_2{\alpha}^2\big)}{\Delta}+  2^{\frac{2}{3}} \Delta\Bigg]
    \label{eqn:23}
\end{equation}
where,
\begin{eqnarray}
 \Delta = \Bigg[ -162A^3 g_2^3-2g_3^3+27Ag_2^2\bigg(-2g_1-6Ag_3+(2-m^2+m^4){\alpha}^2\bigg)\nonumber\\
    +9g_2g_3\bigg(-g_1-4Ag_3+(2-m^2+m^4){\alpha}^2\bigg) +
    \Bigg(\big(54Ag_1g_2^2 + 162A^3g_2^3 \nonumber\\
    + 9g_1g_2g_3+16A^2g_2^2g_3 +36Ag_2g_3^2 +2g_3^3
    -9(2-m^2+m^4)g_2(3Ag_2+g_3){\alpha}^2\big)^2 \nonumber\\
    + 4\big(-(6Ag_2+g_3)^2 
    +3g_2(-g_1+6A^2g_2+ (2-m^2+m^4){\alpha}^2\big)^3\Bigg)^{\frac{1}{2}}\Bigg]
    \label{eqn:20}
\end{eqnarray}

\begin{figure}
    \centering
    \includegraphics[width=0.45\textwidth]{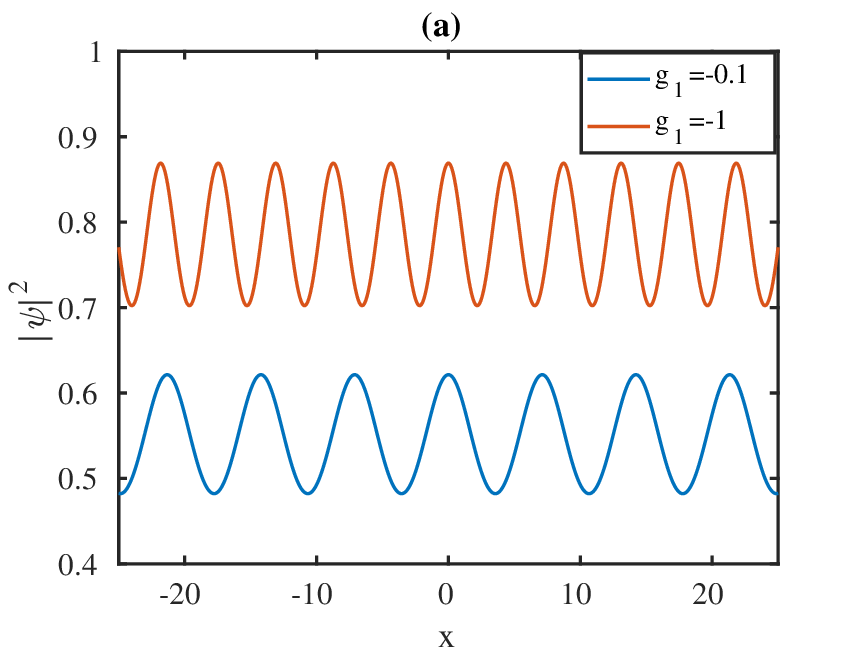}
      \includegraphics[width=0.45\textwidth]{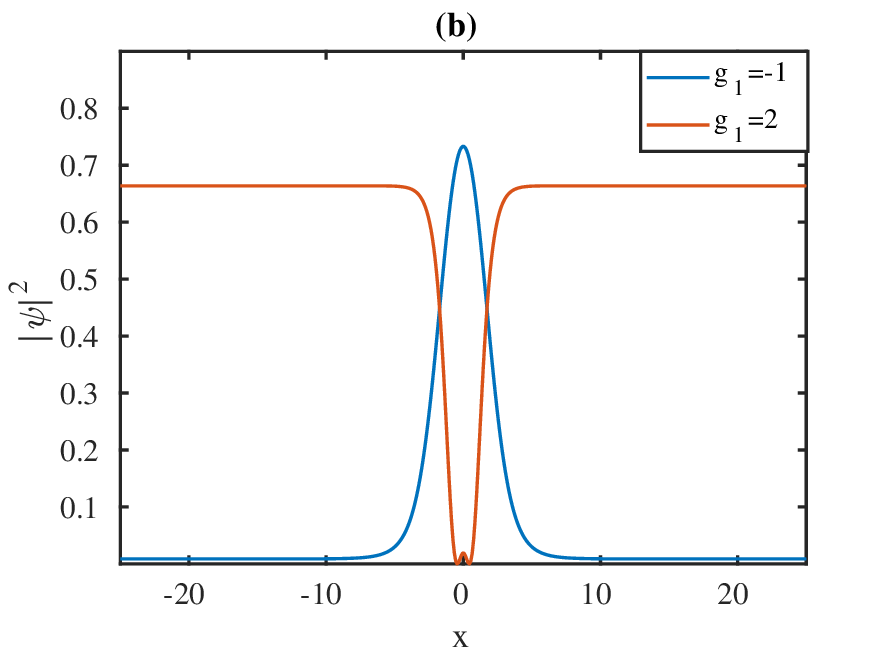}
    \caption{ The density variation of the solution is shown for $g_2=0.3$, $g_3=1$, $M=1=\hbar$ and $g_1$ is mentioned in the legend corresponding to the each plot. Panel (a) represents the profile variation for modulus parameter, $m=0.5$ and Panel (b) depicts for $m=1$, where one can see the variation of profile as a localized bright and dark solitons.}
  \label{fig:2}
\end{figure}

The localized solitons are usually robust, we have made numerical simulations to check the stability of the localized form of the solution i.e., the hyperbolic solution. Figure \ref{fig:2} depicts the density profile of the solution for the different values of modulus parameter. For m=1, a localized bright pulse can be seen for $g_1=-1$.  The localized nature of bright solitons allows them to behave as quasi-particles, maintaining their shape and motion through collisions and interactions with the surrounding medium. This phenomenon is often associated with superfluidity, where the solitons maintain their structure and momentum over long distances, exhibiting persistent currents. The  Gaussian profile of bright solitons signifies a coherent and stable structure with a uniform amplitude distribution over a certain region. A dark pulse can also be obtained for positive value of $g_1$.  Dark pulse, characterized by a localized density dip, exhibit a different energy distribution due to their unique phase properties and interactions with the background medium. The dip profile of dark solitons signifies the localized region of reduced amplitude within the system. This dip represents a phase change, where the soliton exhibits a distinct phase compared to the surrounding medium.  The width of the dip profile is inversely proportional to the interaction strength. Therefore, lower interaction strengths result in broader dips, while higher interaction strengths lead to narrower dips.


In the next sections, we compute the energy as well as the momentum of the solitonic excitations emerging from  the augmented GP equation.

\section{Energy of the soliton}
The Hamiltonian density is given by
\begin{equation}
\mathcal{H}=\frac{{\hbar}^2}{2m} {\Big| \frac{\partial \psi}{\partial x}\Big|}^2 +\frac{1}{2}\delta g |\psi|^4 -\frac{2}{3}\frac{\sqrt{2m}}{\pi \hbar} g^{\frac{3}{2}}|\psi|^3.
\label{eqn:24}
\end{equation}
We consider $\psi(x,t)=\phi(X) e^{[i(kx-\omega t)]}$, with $\phi(X)=\frac{A + B \sech[X]}{1+ D \sech[X]}$ and use the amended NLSE:
\begin{eqnarray}
    \frac{\partial \phi}{\partial t}=-v \frac{\partial \phi}{\partial X}\,\,    \text{ with \,  $v=\frac{\hbar k}{m}$}, \nonumber\\
     \hbar \omega \phi=-\frac{{\hbar}^2{\alpha}^2}{2m}  \frac{\partial^2 \phi}{\partial X^2}+ \frac{{\hbar}^2k^2}{2m}\phi + \delta g |\phi|^2\phi-
    \frac{\sqrt{2m}}{\pi \hbar} g^{\frac{3}{2}}|\phi|\phi.
    \label{eqn:26}
\end{eqnarray}
Considering the case of real $\phi$ and the above can be obtained in the form,

\begin{equation}
    \frac{{\hbar}^2 {\alpha}^2}{2m}  \frac{\partial}{\partial X}\Big[\Big(\frac{\partial \phi}{\partial X}\Big)^2\Big]=-\Big(\mu-\frac{{\hbar}^2k^2}{2m}\Big) \frac{\partial {\phi}^2}{\partial X} + \frac{\delta g}{2}\frac{\partial {\phi}^4}{\partial X}-\frac{1}{3}\frac{\sqrt{2m}}{\pi \hbar} g^{\frac{3}{2}}\frac{\partial {\phi}^3}{\partial X}.
    \label{eqn:27}
\end{equation}
In order to obtain the converged energy, constant background density has to be subtracted from the Hamiltonian. The difference between two asymptotic ends of the soliton is $a=\frac{4 (B-AD)^2}{(1-D^2)^2}$. We now consider the rest-frame with $k=0$ and integrate the above equation with appropriate integration-limit,
\begin{equation}
    \frac{{\hbar}^2 {\alpha}^2}{2m} \int_{X}^{\infty} \frac{\partial}{\partial X}\Big[\Big(\frac{\partial \phi}{\partial X}\Big)^2\Big] dX=-\mu \int_{\phi}^a \frac{\partial {\phi}^2}{\partial X} dX +\frac{\delta g}{2} \int_{\phi}^a \frac{\partial {\phi}^4}{\partial X} dX - \frac{1}{3}\frac{\sqrt{2m}}{\pi \hbar} g^{\frac{3}{2}}\int_{\phi}^a \frac{\partial {\phi}^3}{\partial X} dX,
    \label{eqn:28}
\end{equation}
leading to 
\begin{eqnarray}
    \frac{{\hbar}^2 {\alpha}^2}{2m}\Big(\frac{\partial \phi}{\partial X}\Big)^2&=&-\mu {\phi}^2 + \frac{\delta g}{2}{\phi}^4-\frac{1}{3}\frac{\sqrt{2m}}{\pi \hbar} g^{\frac{3}{2}}{\phi}^3
    \mu \bigg[\frac{4 (B-AD)^2}{(1-D^2)^2}\bigg] \nonumber\\ 
    &-&\frac{\delta g}{2}\bigg[\frac{4 (B-AD)^2}{(1-D^2)^2}\bigg]^2 -\frac{1}{3}\,\frac{\sqrt{2m}}{\pi \hbar} g^{\frac{3}{2}} \bigg[\frac{2 (B-AD)}{(1-D^2)}\bigg]^3.
\label{eqn:29}
\end{eqnarray}
Using Eq. \ref{eqn:29}, the energy with chemical potential ($\mu$) can be represented as:
\begin{eqnarray}
    \mathcal{E}=\int_{-\infty}^{\infty} dX\,  \mathcal{H}&=& \int_{-\infty}^{\infty}  dX \Bigg[\delta g {\phi}^4-\frac{\sqrt{2m}}{\pi \hbar} g^{\frac{3}{2}}{\phi}^3 -2\mu {\phi}^2 +\mu \bigg[\frac{4 (B-AD)^2}{(1-D^2)^2}\bigg] \nonumber\\
    &+& \bigg[\frac{4 (B-AD)^2}{(1-D^2)^2}\bigg]^2 
    - \frac{1}{3}\,\frac{\sqrt{2m}}{\pi \hbar}
    g^{\frac{3}{2}} \bigg[\frac{2 (B-AD)}{(1-D^2)}\bigg]^3\Bigg].
    \label{eqn:31}
\end{eqnarray}

It is important to notice that the last three constant terms are the background energy which compensate the divergent terms in the energy density. After removing the divergent terms, one gets
\begin{eqnarray}
    \mathcal{E}=\delta g \Bigg[ 2 \frac{1}{(1-D^2)^{\frac{7}{2}}} \Big(-6A^2 B^2D(4+D^2)-B^4D(3+2D^2)+4A^3B(2+3D^2)\nonumber\\ +4AB^3(1+4D^2)
    +A^4D(-8+8D^2-7D^4+2D^6\Big)tan^{-1}\Big(\sqrt{\frac{1-D}{1+D}}\Big)\Bigg] \nonumber\\ +2\mu 
    \Bigg[\frac{4}{(1-D^2)^{\frac{3}{2}}}(B-AD)\Big(BD+ A(-2+D^2)\Big)tan^{-1}\Big(\sqrt{\frac{1-D}{1+D}}\Big)\Bigg] \nonumber\\ + \frac{\sqrt{2m}}{\pi \hbar} g^{\frac{3}{2}} \Bigg[ \frac{2}{(1-D^2)^{\frac{5}{2}}}(-B+AD) +
    \Big(2ABD(-4+D^2)+B^2(1+2D^2) \nonumber\\
    + A^2(6-5D^2+2D^4)\Big)tan^{-1}\Big(\sqrt{\frac{1-D}{1+D}}\Big)\Bigg]. 
    \label{eqn:44}
\end{eqnarray}

\section{Momentum of the soliton}
The momentum expession is given by
 \begin{equation}
    \mathcal{ P}=-\frac{\iota \hbar}{2}\int \big[\psi^{*} \frac{\partial \psi}{\partial x}- \psi \frac{\partial \psi^{*}}{\partial x}\big]dx.
     \label{eqn:45}
 \end{equation}
 We consider the solitonic solution  with a finite length $L$ with interval $[- L/2, L/2]$ and take the limit $L \longrightarrow \infty$, into the form
 $\psi(x,t)=\frac{A+ B sech(X) }{1+ D sech(X)}e^{[i(kx-\omega t)]}$:
 \begin{equation}
 \begin{aligned}
     \mathcal{ P}&=-\frac{i\hbar}{2}\int_{-\frac{L}{2}}^{\frac{L}{2}} \big[\psi^{*} \frac{\partial \psi}{\partial x}- \psi \frac{\partial \psi^{*}}{\partial x}\big]dx\\
     &= k \hbar\Bigg[ A^2 \alpha L + 2 (B-AD) (BD+ A(-2+D^2))2 tan^{-1}\bigg(\frac{D-1}{\sqrt{D^2-1}}\tanh(\frac{\alpha L}{4})\bigg)\Bigg]
     \label{eqn:47}.
     \end{aligned}
 \end{equation}
Here we can correlate the first-term to the background $e^{[i(kx-\frac{\mu t}{\hbar})]}$ and  the second term  comes from usual solitonic contribution. In the appropriate limit case the momentum yields into 
\begin{equation}
    \mathcal{P}=4 k\hbar (B-AD) (BD+ A(-2+D^2))tan^{-1}\bigg(\frac{D-1}{\sqrt{D^2-1}}\bigg).
    \label{eqn:48}
\end{equation}
\section{Conclusion}
In conclusion, We have systematically investigated the solution space of the augmented NLSE through a  fractional transformation method, incorporating the Jacobi elliptic function \text{dn}[x, $m^2$]. It analyses the exact solution in the form of droplets and supersolid. By accurately determining the momentum and energy of the soliton, we gain valuable insights into the underlying physical principles governing the system's dynamics. The momentum characterizes the soliton's translational motion and provides information about its velocity and direction. On the other hand, the energy quantifies the soliton's total energy content, including both its kinetic and potential energy contributions.

The obtained results demonstrate the emergence of a cigar-type bright soliton, illustrating the formation of a stable, coherent, and localized structures within the system. On the other hand, dark solitons, show a distinctive energy distribution because of their distinct phase features and interactions with the background medium. They are distinguished by a localized density dip. This soliton behavior is a hallmark of the superfluid nature of the system, where the quantum particles exhibit a remarkable collective coherence and flow without dissipation.
Understanding the energy and momentum properties of solitons is of great importance for a range of applications. In superfluid systems, solitons can be manipulated and controlled to transport energy and momentum, making them potential candidates for information transfer and storage in quantum communication and computing devices. Furthermore, the study of soliton dynamics in superfluid systems contributes to our understanding of various physical phenomena, such as Bose-Einstein condensates and topological excitations.

Looking ahead, future research efforts may involve investigating the effects of external potentials or interactions between multiple solitons on the superfluid behavior and energy-momentum characteristics. Additionally, exploring the impact of dissipation and thermal effects on soliton dynamics would provide valuable insights into the robustness and stability of these systems.

\newpage

\end{document}